\documentclass[reprint,preprintnumbers,amsmath,amssymb,nofootinbib,aps,prx]{revtex4-2}

\usepackage{slashed}
\usepackage{amsmath}
\usepackage{amssymb}
\usepackage{amsthm}
\usepackage[dvipsnames]{xcolor}
\usepackage{hyperref}
\usepackage{indentfirst}
\usepackage{psfrag}
\usepackage{graphicx}
\usepackage{cleveref}
\Crefname{equation}{Eq.}{Eqs.}
\usepackage[utf8]{inputenc}

\hypersetup{colorlinks=true, linkcolor=blue, urlcolor=blue, citecolor=blue, linktocpage=true}

\newcommand{\cm}{\ensuremath{\mathrm{cm}}}

\newcommand{\eV}{\ensuremath{\mathrm{eV}}}

\newcommand{\MeV}{\ensuremath{\mathrm{MeV}}}

\newcommand{\E}{\mathcal{E}}

\newcommand{\D}{\ensuremath {\mathrm{d}}}

\newcommand{\ane}{\bar{\nu}_e}
\renewcommand{\ne}{\nu_e}
\newcommand{\anm}{\bar{\nu}_\mu}
\newcommand{\ant}{\bar{\nu}_\tau}
\newcommand{\nm}{\nu_\mu}
\newcommand{\nt}{\nu_\tau}
\newcommand{\LL}{L_\mu-L_\tau}
\newcommand{\LLo}{L_e-L_{\mu/\tau}}

\renewcommand{\d}{\partial}
\newcommand{\g}{\gamma}

\renewcommand{\L}{\mathcal{L}}
\newcommand{\beq}{\begin{equation}}
\newcommand{\eeq}{\end{equation}}
\newcommand{\bea}{\begin{eqnarray}}
\newcommand{\eea}{\end{eqnarray}}

\newcommand*\diff{\mathrm{d}}
\let\limitint\int % Only when I provide explicit limits for the integration, I need to do the spacing myself
\renewcommand{\int}{\limitint \!} % The standard integral should have correct spacing
\newcommand\e{\text{e}}

\newcommand{\vect}[1]{\boldsymbol{#1}}

\begin{document}

\vspace{-3.0cm}
\begin{flushright}
{\small UMN-TH-4530/26} 
\end{flushright}
\vspace{0.5cm}

\title{New Gauge Forces, Neutron Stars and Schwinger Neutrino Production}

\author{Yuxin Liu$^1$}
\email{liuyuxin211@mails.ucas.ac.cn}
\author{Zhen Liu$^2$}
\email{zliuphys@umn.edu}
\author{Andrey Shkerin$^3$}
\email{ashkerin@perimeterinstitute.ca}
\author{Jing Shu$^{4,5,6}$}
\email{jshu@pku.edu.cn}
\author{Yue Zhao$^7$}
\email{zhaoyue@ust.hk}

\affiliation{$^1$International Center for Theoretical Physics Asia-Pacific (ICTP-AP), University of Chinese Academy of Sciences (UCAS), Beijing 100190, China}
\affiliation{$^2$School of Physics and Astronomy, University of Minnesota, Minneapolis, Minnesota 55455, USA }
\affiliation{$^3$Perimeter Institute for Theoretical Physics, 31 Caroline St N, Waterloo, ON N2L 2Y5, Canada}
\affiliation{$^4$School of Physics and State Key Laboratory of Nuclear Physics and Technology, Peking University, Beijing 100871, China}
\affiliation{$^5$Center for High Energy Physics, Peking University, Beijing 100871, China}
\affiliation{$^6$Beijing Laser Acceleration Innovation Center, Huairou, Beijing, 101400, China}
\affiliation{$^7$Department of Physics and Jockey Club Institute for Advanced Study, The Hong Kong University of Science and Technology, Hong Kong S.A.R., P.R.China}

\begin{abstract}

We investigate neutrino effects of new long-range forces arising from gauging $B-L$, $\LLo$ or $\LL$ symmetries of the Standard Model. The leptonic potential generated by astronomical bodies, such as the Earth, the Sun or a neutron star, results in the Schwinger pair production of neutrinos charged under the new gauge symmetry. The oppositely charged particles accumulate in the potential well forming a degenerate Fermi gas, while equally charged particles fly away forming a steady flux of neutrinos. We find that, for the $B-L$ and $\LLo$ forces, these effects are too weak to be observable. For the $\LL$ force these effects are significant in neutron stars if the gauge coupling is $g\gtrsim 10^{-18}$. 
The muonic force changes the element abundances of a neutron star in equilibrium and suppresses its $\LL$ charge. This \textit{invalidates} the constraint on $g$ from neutron star mergers, at $g\gtrsim 10^{-17}$.
Furthermore, for such values of $g$, the neutrino flux produced by the Schwinger effect could potentially be detected from a single young neutron star at a distance of $\simeq 100$ pc, with the typical neutrino energy $E_\nu\sim 100$ MeV.
A dedicated search for such a signal will reassert the bound $g\lesssim 10^{-18}$.

\end{abstract}

\maketitle

\section{Introduction}
\label{Intro}

Extensions of the Standard Model containing light vector particles are a subject of active research.
New light $U(1)$ gauge bosons give rise to a variety of observational signatures \cite{ParticleDataGroup:2022pth}. 
They mediate long-range forces that are constrained by experiments looking for a fifth force or equivalence principle violation; see \cite{Antypas:2022asj} for a review.

Consider the new gauge boson $A_\mu$ of mass $m_A$ coupled directly to the Standard Model (SM) matter fields,
\beq\label{L1}
 \L_{int.} = g A_\alpha J^\alpha_{SM} \;,
\eeq
where the current $J^\alpha_{SM}$ is associated with one of the global SM charges.
Cancellation of anomalies within the SM limits the linearly independent conserved currents to $B-L$, $L_e-L_\mu$, $L_e-L_\tau$ or $\LL$.  The leptonic part of the current $J^\alpha_{SM}$ contains active neutrinos. For example, for the $L_a-L_b$ current, the new gauge boson-neutrino interaction reads
\beq \label{L2}
\L_{int.}=g A_\alpha ( \bar{\nu}_a\g^\alpha\nu_a - \bar{\nu}_b\g^\alpha\nu_b ) \;.
\eeq
In this paper, we will explore phenomenological consequences of this interaction.

A theory of the massive gauge boson is not complete and must be supplemented, e.g. with the Higgs mechanism.
Furthermore, the gauge $L_a-L_b$  symmetry is broken explicitly by the active neutrino masses.
It is natural to expect that the same mechanism gives masses to the neutrinos as well~\cite{Ma:2001md,Choubey:2004hn,Heeck:2011wj,Asai:2018ocx}.
Observational constraints on the new gauge coupling $g$ and mass $m_A$ can be insensitive to the details of the completion~\cite{Dror:2020fbh}, and
we will assume this to be the case in this work.

Large baryon and electron charges of astronomical bodies, such as the Earth or the Sun, lead to a large new gauge boson field strength sourced by the charge, amplifying effects arising from coupling to the SM.
As a result, the gauge couplings of $B-L$ and $\LLo$ currents are severely constrained by the fifth force experiments \cite{Adelberger:2003zx,Adelberger:2006dh,Schlamminger:2007ht}.
These constraints are not applicable to the $\LL$ current, since ordinary matter does not contain muons and tauons.

Recently it has been realized that neutron stars are a powerful probe of dark sectors; see \cite{Gardner:2023wyl} for a review. For example, 
dark matter can impact formation, structure and evolution of a neutron star (NS) and affect orbital periods of NS binaries \cite{Blas:2016ddr,Armaleo:2019gil,KumarPoddar:2019ceq,Dror:2019uea}.
Neutron stars also stand out in that they contain a significant ($\sim 0.1-1\%$) mass fraction of muons \cite{haensel2007neutron}. 
Hence they are natural laboratories to study dark sectors coupled to the second generation of fermions \cite{Garani:2018kkd,Bell:2019pyc,Garani:2019fpa,KumarPoddar:2019ceq,Dror:2019uea}.
In particular, Refs.~\cite{KumarPoddar:2019ceq,Dror:2019uea} studied how the new gauge boson coupling to the $\LL$ current creates both an additional force and dark radiation in NS binaries, putting strong constraints on the gauge coupling for $m_A$ below the $10^{-10}\,$eV scale.

When the gauge coupling $g$ is sufficiently large, the muonic component of the $\LL$ force shifts the balance of chemical reactions involving leptons and alters the equation of state of a NS. As a result, the equilibrium muon number and the total muon charge of the NS {may} deviate from their values at $g=0$.
In this paper, we calculate this back-reaction effect by solving for the NS equation of state with the new force taken into account.\footnote{We assume that the impact of the new force on the nuclear part of the equation of state is negligible.} 
We find that the equilibrium NS muonic charge is affected by the new force at $g\gtrsim 10^{-18}$, in agreement with the discussion in Ref.~\cite{Dror:2019uea}.
The effect becomes stronger with the stronger coupling, and at $g \gtrsim 10^{-17}$ (and gauge boson masses $m_A\lesssim R_{NS}^{-1}\sim 10^{-10}$ eV) the bound on $g$ from NS binaries calculated in \cite{Dror:2019uea} is not applicable.

Neutrinos, as well as muons, experience the presence of the $\LL$ force.
Muon (anti)neutrinos play the crucial role during the NS birth and cooling. Besides, they participate in maintaining the chemical equilibrium of the NS \cite{haensel2007neutron}. 
Normally, they do not accumulate enough in the NS to affect the balance of weak reactions in equilibrium. 
This is because the energy of produced neutrino is determined by the NS temperature, $T_{\rm NS}\lesssim 10$ MeV~\cite{haensel2007neutron},
and they eventually escape the NS interior. We will see that the $\LL$ electric potential generated by the new force can be strong enough to confine muon antineutrinos, which screen the muonic charge.
Note that the confined neutrino cannot escape by oscillating into a different flavour, since neutrino oscillations are suppressed in the presence of strong $\LL$ field.

Yet another consequence of the interaction (\ref{L2}) is that neutrino-antineutrino pairs can be spontaneously produced in strong enough new gauge fields, analogously to the Schwinger electron-positron pair production in strong electromagnetic fields. Feasible production rate requires the flavor neutrino mass be sufficiently small. One of the particles of the produced pair is attracted by the potential. Thus, astronomical bodies, such as the Earth or the Sun, could potentially host ``Fermi-balls'' of neutrino particles. We will see that for the $B-L$ and $\LLo$ forces the existing constraints on the gauge couplings preclude this effect from being observable.

However, the situation is different for the $\LL$ force. The Schwinger production of neutrinos in the NS potential leads to two effects. First, the Fermi-balls of muonic antineutrinos and tau neutrinos change the NS equation of state and significantly screen the NS charge. 
Second, before this equilibrium state is reached, the NS emits the flux of muonic neutrinos and tau antineutrinos. We will see that 
the flux is peaked at the neutrino energy $E_\nu\sim 100$ MeV,
and the typical NS ``discharge'' time is $t_{Q}\gtrsim 10^7$ yr. 
We estimate that a single young NS located 100 pc away from the Earth generates the Schwinger neutrino flux that at its peak energy matches or even exceeds the atmospheric neutrino flux in the direction of the NS.
Non-observation of this additional flux in a directional neutrino search will exclude the values $g\gtrsim 10^{-18}$.

The paper is organised as follows. In Sec.~\ref{sec:neutr} we discuss the Schwinger effect for neutrinos and make estimates for the effect from $B-L$ and $\LLo$ forces generated by the Earth or the Sun. We also make preliminary estimates of the effect of the $\LL$ force on a NS and of the generated neutrino flux. In Sec.~\ref{sec:NS} we discuss the equation of state of the NS in the presence of the $\LL$ force. We calculate the impact of the muonic potential on the equilibrium number densities of muons and muon antineutrinos\footnote{Tau neutrinos produced, e.g. by the Schwinger mechanism, are also present in the equilibrium state.}
and find how the muon number depends on the gauge coupling in equilibrium. 
We also outline the main stages of the approach to the equilibrium state. In Sec.~\ref{sec:pheno} we discuss the two main consequences of our analysis: opening of the window $g\gtrsim 10^{-17}$ where the NS binaries bound is not applicable, and a new prospective bound $g\lesssim 10^{-18}$ using the Schwinger neutrino flux from a young nearby NS.
We conclude in Sec.~\ref{sec:concl}.
Several Appendices contain details of the calculations.

\section{Neutrino Schwinger effect}
\label{sec:neutr}

Spontaneous $e^+e^-$-pair creation in strong electric fields \cite{sauter1931verhalten,heisenberg2006consequences,Schwinger:1951nm} is a well-studied nonperturbative quantum effect \cite{Schwartz:2014sze}.
We are interested in physical implications of this effect in the case when the electric field is due to one of the new gauge forces.
We will study the neutrino pair creation, since other leptons are too heavy to be produced by this mechanism.
Consider an active neutrino of flavour $a$ coupled to the $B-L$ or $L_a-L_b$ gauge field as per \Cref{L2}.
The production rate of $\nu_a\bar{\nu}_a$ pairs by a constant electric field of strength $\E$ is given by
\beq \label{SchwRate}
\Gamma_{\E\to\nu_a\bar{\nu}_a} = \frac{  g^2\E^2}{4\pi^3} \exp\left[ - \frac{\pi m_a^2}{g\mathcal{E}} \right] \;.
\eeq
Here $m_a$ is the ``effective mass'' of the neutrino of flavour $a$, which is related to the mass eigenstates as $m_a=\sum_i m_i |V_{ai}|^2$, where $V_{ai}$ is the $3\times 3$ active neutrino mixing matrix in medium and in the presence of the new force.
If $\E$ is space-dependent, the expression (\ref{SchwRate}) determines the local production rate as long as the spatial variation scale of $\E$ is much bigger than the de Broglie wavelength of the produced neutrino.\footnote{One can estimate the de Broglie wavelength from the uncertainty relation: $\Delta x \sim (\Delta p)^{-1}$. Using that $\Delta p\sim\sqrt{g\E}$ (see \cite{Cohen:2008wz} and Sec.~\ref{sec:pheno}), we obtain $\Delta x\sim 1/\sqrt{g\E}$.}

We now make a crucial assumption that the flavour mass $m_a$ is small enough to overcome the exponential suppression factor, 
\beq \label{ma1}
m_a\lesssim \sqrt{g\E/\pi} \;,
\eeq
so that the rate is given by the pre-exponential part of \Cref{SchwRate}. While there is no lower experimental bound on any of $m_a$ \cite{ParticleDataGroup:2024cfk}, for the most interesting case of the $\LL$ force in the NS, this condition gives $m_{\nu_{\mu/\tau}}\lesssim 0.05$ eV. For simplicity, below we assume that $m_{\nu_\mu}\approx m_{\nu_\tau}$.

Consider a macroscopic source of the new gauge force, such as the Earth, the Sun or an NS. We would like to see if the neutrino pair creation process can possibly lead to observable effects. {We envisage} two types of effects. First, one particle from the created pair (antineutrino $\bar{\nu}_a$) is trapped in the potential well. Over time, accumulated $\bar{\nu}_a$-particles partially screen the charge. Their number density is limited {theoretically} by the Pauli blocking. Second, the other particle (neutrino $\nu_a$) is repelled from the source forming a steady flux of typical energy $E_\nu\sim-g\phi_{max}$ where $\phi_{max}$ is the depth of the potential well.

Below we first estimate the screening effect of the Schwinger-produced neutrino in the $B-L$ or $\LLo$ potential of the Earth or the Sun. 
The crude estimate assumes that all produced (anti)neutrinos are trapped in the potential well, where they can relax and form the Fermi--Dirac distribution.
The relaxation can indeed happen due to scattering on matter or due to neutrino Bremsstrahlung in the $B-L$ or $\LLo$ potential, if the gauge coupling is large enough.
However, we find that the screening effect is too weak to change the existing bounds on the gauge couplings, and the latter are too strong for the Schwinger effect to be observable.
We then turn to the $\LL$ potential of a NS, for which we find that both effects can be important. 
Furthermore, the trapped (anti)neutrinos quickly thermalize to the NS temperature, and they follow the Fermi--Dirac distribution to a good approximation.\footnote{The mean free path of neutrino with the energy $E_\nu = 10-100\:\MeV$ in a NS is short compared to the NS size; see e.g. \cite{Reddy:1997yr}.
Furthermore, in the $\LL$ electric field, neutrinos can spontaneously transition to unoccupied lower-energy levels.
The standard electric-dipole estimate is $\Gamma_{\rm dip}\sim g^2(\Delta E)^3R^2/(4\pi)$, where $\Delta E$ is the energy difference between the levels and $R$ is the coherent size of the radiating charge distribution. Taking, for example, $\Delta E\sim 10$ MeV and $R\sim\Delta E^{-1}\ll R_{NS}$, we obtain $\Gamma_{\rm dip}^{-1}\sim (10^{-18}/g)^2\cdot 10^5$ yr, which is much smaller than the discharge time; see \Cref{tQ} and Sec.~\ref{ssec:sn}.}
Motivated by this result, in Sec.~\ref{sec:NS} we study the NS case in detail.

\subsection{$B-L$ or $\LLo$ force}
\label{ssec:b-l}

Consider the Earth as a source of the $B-L$ or $\LLo$ force. We approximate the Earth by a static, spherically symmetric and uniformly charged body with the parameters
\beq \label{numbers}
n\approx 1.7\cdot 10^{24} \:\cm^{-3}\;, ~~ R\approx 6400\:{\rm km} \;,
\eeq
where $n$ is an approximate average number density of electrons or neutrons, $R$ is the Earth radius.
Assume also that the new gauge boson is light enough, $m_A\lesssim 10^{-14}\:\eV\sim R^{-1}$; otherwise the potential due to the new force is suppressed by a factor $1/(m_AR)$; {see Appendix~\ref{app:A}.}
At $r<R$ the potential and the field strength read as follows,
\beq\label{phiE}
    \phi(r)=-gn\frac{3R^2-r^2}{6} \;,~~ \E(r)=-\phi'(r)=\frac{gnr}{3} \;,
\eeq
where prime means derivative with respect to $r$. The maximum density of antineutrino in the potential well is given by the Fermi density 
\beq
    n_F(r)=8\pi g^3|\phi(r)|^3 \;,
\eeq
 and the total number of (anti)neutrinos that can occupy the potential well is, per flavor,
\beq\label{N}
    N=4\pi \limitint_0^R n_F(r)r^2\diff r=\frac{6016}{8505}\pi^2g^6 n^3 R^9 \;.
\eeq
Let us see if this number is enough to screen the $B-L$ or $\LLo$ charge of the Earth at the distance $r\approx R$, where the fifth-force experiments are performed. {The charge is $Q=4\pi R^3n/3$. Using \Cref{N} and substituting the numbers (\ref{numbers}), we obtain}
\beq \label{NQ}
    \frac{N}{Q}\sim ( 10^{16}g)^6 \;.
\eeq
Thus, to have the significant screening effect, we need the coupling $g\gtrsim 10^{-16}$.
These values are, however, excluded by short-range tests of the equivalence principle~\cite{Smith:1999cr}, which give $g\lesssim 10^{-22}$ for $g_{B-L}$ and $g_{e-\mu/\tau}$, and do not rely on the macroscopic charge of the Earth.
We conclude that the neutrino Schwinger pair production by the $B-L$ or $\LLo$ potential of the Earth is too weak to be observable. Repeating the above estimates for the Sun, we arrive at the same conclusion.

\subsection{$\LL$ force}
\label{ssec:NS}

For NSs, the situation is different: a significant fraction of the NS matter is in the form of muons, opening the possibility to test the muonic, $\LL$ force, which is not tightly constrained by fifth-force experiments. In the estimates below we take $g=10^{-18}$ as a benchmark value of the gauge coupling \cite{Dror:2019uea}. We also approximate a NS to be a static, spherically symmetric and uniformly charged body, leaving the more realistic analysis to the next section. We take
\beq \label{numbers2}
n\approx 5\cdot 10^{37}\:\cm^{-3} \;, ~~ R\approx 10\:{\rm km} \;,
\eeq
as the number density of muons, which corresponds to $1\%$ mass fraction of muons in a star with $M=2M_{\odot}$, and the NS radius, correspondingly~\cite{Ozel:2016oaf}. Using \Cref{phiE}, we estimate the potential energy generated by muons:
\beq \label{phiMax}
g|\phi_{\rm max}|=\frac{1}{2}g^2nR^2 \sim 100\:\text{MeV}\:\left(\frac{g}{10^{-18}}\right)^2 \;.
\eeq
With this benchmark value, the potential energy is comparable to the muon mass. Thus, at $g\gtrsim 10^{-18}$ the muonic component of the $\LL$ force can significantly impact the NS state and equilibrium abundances of leptons \cite{Dror:2019uea}. 
From \Cref{phiMax} we can also estimate the typical energy of the (anti)neutrino generated by the Schwinger effect: $E_\nu\sim g|\phi_{\rm max}|\sim 100$ MeV for $g\sim10^{-18}$. Next, using \Cref{NtSmall,numbers2} we estimate the neutrino flux at a distance $L$ from the NS, per flavor:
\beq \label{NuFlux}
\frac{\diff N_t}{\diff t} \frac{1}{4\pi L^2}\sim 10^{-2}\left( \frac{g}{10^{-18}}\right)^4 \left(\frac{100\:\text{pc}}{L}\right)^2 \frac{1}{\text{cm}^2\cdot\text{s}} \;.
\eeq
This is comparable to the atmospheric neutrino flux \cite{Gaisser:2002jj} at $E_\nu\sim 100$ MeV and $g\sim 10^{-18}$ provided that we have a good directional resolution of the incoming neutrino; see Fig.~\ref{fig:Neutrino_Flux} below. 

However, the flux diminishes with time due to the charge screening by accumulated muon antineutrinos and tau neutrinos. 
Let us estimate the back-reaction time.
The unscreened charge number density is given by
\beq\label{nEff}
    n_{\rm eff}(r,t)=n-n_\nu(r,t) \;,
\eeq
where $n_{\nu}$ is the number density of (anti)neutrino trapped in the potential by time $t$. 
Replacing $n$ by $n_{\rm eff}(r,t)$ in \Cref{phiE} and using \Cref{SchwRate}, we obtain
\beq
      \frac{dn_\nu(r,t)}{dt}  = \Gamma(r,t) = \frac{g^4(n-n_\nu(r,t))^2r^2}{36\pi^3} \;.
\eeq
Solving this equation with the initial condition $n_\nu(r,0)=0$ gives 
\beq
    n_\nu(r,t)=n \left( 1 - \frac{1}{1+\frac{g^4nr^2t}{36\pi^3}} \right) \;.
\eeq
From here we find the total number of (anti)neutrino created by time $t$:
\beq\label{Nt}
\begin{split}
    N_t & =4\pi\limitint_0^R n_\nu(r,t)r^2\diff r\\
    & =4\pi n\left[ \frac{R^3}{3}-\frac{R}{at}+\frac{\arctan(\sqrt{at}R)}{(at)^{3/2}} \right] \;,
\end{split}
\eeq
where $a=g^4n/(36\pi^3)$. At small $at$, when the back-reaction of (anti)neutrino on the potential is small, \Cref{Nt} gives
\beq\label{NtSmall}
    N_t\approx \frac{g^4n^2R^5t}{45\pi^2} \;.
\eeq
From here we find the time needed to reach order-one screening:
\beq\label{tQEarth}
t_Q\approx \frac{Q}{\diff N_t/\diff t} = \frac{60\pi^3}{g^4nR^2} \;.
\eeq
Plugging in the numbers (\ref{numbers2}), we obtain
\beq \label{tQ}
t_Q\approx \left( \frac{10^{-18}}{g} \right)^4 3\cdot 10^7\: {\rm yr.}
\eeq
Thus, only sufficiently young NSs, not yet discharged by the Schwinger mechanism, emit the neutrino flux. 

These estimates show that the $\LL$ force with $g\gtrsim 10^{-18}$ leads to two phenomenologically interesting effects. First, it modifies the NS equation of state, necessitating the reassessment of the bound on the gauge coupling in the region $g\gtrsim 10^{-18}$, since the existing bound relies on the muonic mass fraction of a NS being unaffected by the new force.
Second, non-observation of the cosmic neutrino flux from NSs due to the Schwinger pair production imposes a novel prospective bound on $g$. 

The above estimates need to be improved in several ways. The distribution of muons in a NS is not uniform, the spacetime around it is not flat. Even more importantly, the abundance of muons (and other species) must be determined consistently by the conditions of chemical and hydrostatic equilibrium and in the presence of the new force. The estimate of the muonic potential energy (\ref{phiMax}) shows that the $\LL$ force cannot be ignored in determining the NS equilibrium state.
In the remainder of the paper we will calculate the modified equilibrium state and get more accurate estimates of the $\LL$ charge screening and Schwinger neutrino flux.

Let us evaluate the neutrino flavor mass required for the unsuppressed Schwinger production. Using \Cref{ma1,phiE,numbers2}, we obtain
\beq \label{ma2}
m_a\lesssim \frac{g}{10^{-18}}\: 0.05\: \text{eV} \;.
\eeq
As soon as $g\gtrsim 10^{-18}$, this bound is consistent with the cosmological constraint on the sum of neutrino masses $\sum_\nu m_\nu < 0.12$ eV \cite{Planck:2018vyg}.

\section{Neutron star in the presence of $\LL$ force}
\label{sec:NS}

\subsection{General remarks}

In this section we first derive how the $\LL$ force impacts the NS composition in thermal, chemical and mechanical equilibrium. We then discuss how the equilibrium state can be achieved following a progenitor supernova explosion. Finally, we calculate the neutrino flux emitted by the NS due to the Schwinger pair production during the equilibration. 

The analysis requires the knowledge of the NS equation of state.
The latter is not known precisely, and the composition of innermost layers of the NS is subject to debates; see \cite{Page:2006ud,Lattimer:2021emm} for reviews.
We will conservatively assume that the NS consists of neutrons $n$, protons $p$, electrons $e$ and muons $\mu$, which are kept in chemical equilibrium by weak reactions.
We adopt the nuclear equation of state determining the chemical potentials of $n$ and $p$ from Refs.~\cite{Akmal:1998cf,Heiselberg:1999mq}.

We consider an isolated and non-rotating NS, unperturbed by external forces.\footnote{The NS rotation would not significantly change our results, since the resulting $\LL$ magnetic field is suppressed compared to the electric field by the NS velocity $v\ll c$.}
In the presence of the $\LL$ force, the NS acquires a charge due to the abundance of muons and lack of anti-muons.
The charge generates the potential well confining particles of opposite charge that do not have enough energy to escape.
As estimated in Sec.~\ref{sec:neutr}, the depth of the well can reach $\sim 100$ MeV. 
The important consequence is that muon anti-neutrinos $\anm$ produced in the weak reactions with average energy $E_\nu\lesssim 10\:\MeV$ \cite{Li:2023ulf} are partially trapped in the NS interior. The trapped anti-neutrinos quickly thermalise and form the Fermi--Dirac distribution like electrons and muons. More $\anm$ and tau neutrinos $\nt$ produced by the Schwinger mechanism are also trapped.\footnote{
We will neglect tau (anti)neutrinos produced by other mechanisms, since they do not affect the beta-equilibrium. 
} 
The trapped particles partially screen the $\LL$ charge of the NS. The process can continue until the production of $\anm$ and $\nt$ stops due to the Pauli-blocking. This is the new equilibrium state. We would like to calculate the equilibrium number densities $n_n$, $n_p$, $n_e$, $n_\mu$, $n_{\anm}$ and $n_{\nt}$ and the resulting $\LL$ charge.

\subsection{Equilibrium state}
\label{ssec:eos}

\subsubsection{Chemical equilibrium}

The main reactions establishing the chemical equilibrium of a NS are the weak reactions:
\begin{align}
& n \rightarrow p + e + \ane \;, ~~~ p+e \rightarrow n + \ne \label{n_pe} \;, \\
& n \rightarrow p + \mu + \anm \;, ~~~ p + \mu \rightarrow n + \nm \;. \label{n_pmu}
\end{align}
They determine the equilibrium number densities of $n$, $p$, $e$ and $\mu$.
In beta-equilibrium, the chemical potentials of the species satisfy
\beq\label{Mu_npemu}
\mu_n - \mu_p = \mu_e = \mu_{\mu} + \mu_{\anm} \;.
\eeq
In the second equality in \Cref{Mu_npemu} we take into account the contribution from muon anti-neutrino, since the latter can be confined in the $\LL$ potential of the NS.
The electron chemical potential is determined by its Fermi energy:
\beq \label{Mu_e}
\mu_e \approx (3\pi^2 n_\e)^{1/3} \approx 122.1 \left( \frac{n_e}{0.05 n_0} \right)^{1/3}\:\MeV \;,
\eeq
where $n_0=0.15$ fm$^{-3}$ is the nuclear density and we neglected the electron mass.
Here and below we assume that the NS temperature is small enough and one can neglect the temperature-dependent corrections to particle's chemical potentials.
This approximation is very well satisfied at times when the equilibrium state is approached.
The chemical potentials of muonic species receive contributions both due to the Pauli blocking ($\mu^{(k)}$) and the $\LL$ potential ($\mu^{(p)}$).
The first ones read
\beq \label{Mu_mu}
\mu_\mu^{(k)} = \sqrt{m_\mu^2+(3\pi^2 n_\mu )^{2/3}} \;, ~~~\mu_{\anm}^{(k)} = (3\pi^2 n_{\anm})^{1/3} \;,
\eeq
where we take into account the muon mass $m_\mu$ since, unlike electrons, muons are only mildly relativistic.
The chemical potentials $\mu^{(p)}$ and $\mu_{\anm}^{(p)}$ are discussed later.
Finally, one relation between $n_p$, $n_e$ and $n_\mu$ follows from the local SM electric charge neutrality of the NS matter:
\beq\label{neutr} 
n_p = n_e + n_\mu \;.
\eeq

To find $\mu_n$, $\mu_p$ as functions of $n_n$, $n_p$, we need a nuclear equation of state.
The latter relates the difference $\mu_n-\mu_p$ to the nuclear symmetry energy $S(n_b)$, where $n_b$ is the baryon number density, and the asymmetry factor $\delta=(n_n-n_p)/n_b$.
The relation reads as follows,
\beq  \label{Mu_np}
\mu_n - \mu_p = 4 \delta S(n_b) + m_n - m_p \;, ~~~ n_b=n_n+n_p \;,
\eeq
where $m_n$ ($m_p$) is the neutron (proton) mass; see Appendix~\ref{app:B} for details.
We will use the following phenomenological dependence extrapolating the symmetry energy to supra-nuclear densities \cite{Akmal:1998cf,Heiselberg:1999mq,Haensel:2007yy,haensel2007neutron}:
\beq  \label{S_nb}
S(n_b)=32 n_b^{0.6} \: \MeV \;.
\eeq

\subsubsection{Hydrostatic equilibrium}

The starting point is the expression for a static, spherically-symmetric spacetime,
\beq \label{metric}
\diff s^2 = \e^{2\Phi}\diff t^2 -\e^{2\lambda}\diff r^2 - r^2\diff\Omega_2^2 \;,
\eeq
where $\Phi$, $\lambda$ are functions of $r$ and $\diff\Omega_2^2$ is the line element of a unit 2-sphere.
The gravitational mass $m(r)$ confined inside a sphere with radius $r$ is found from the relation
\beq \label{m(r)}
\e^{-\lambda} = \sqrt{1-2Gm/r} \;.
\eeq
Substituting the ansatz (\ref{metric}), (\ref{m(r)}) to the Einstein equations and adopting the perfect fluid approximation of the NS matter, one arrives at the following equations of hydrostatic equilibrium \cite{haensel2007neutron},
\beq\label{hydro}
\begin{aligned}
\frac{\diff P}{\diff r} & = -\frac{G\rho m}{r^2}\left(1+\frac{P}{\rho}\right)\left(1+\frac{4\pi P r^3}{m} \right)\left( 1-\frac{2Gm}{r} \right)^{-1} , \\
\frac{\diff m}{\diff r} & = 4\pi r^2\rho \;, \\
\frac{\diff\Phi}{\diff r} & = -\frac{1}{\rho}\frac{\diff P}{\diff r}\left(1+\frac{P}{\rho} \right)^{-1} \;,
\end{aligned}
\eeq
where $P$ and $\rho$ denote the (space-dependent) pressure and energy density, correspondingly.
These equations must be supplemented with the NS equation of state relating $P$ and $\rho$:
\beq \label{P(rho)}
P = \sum_i \left( n_b\frac{\diff \rho_i}{\diff n_b} - \rho_i \right) \;, ~~ \rho=\sum_i\rho_i \;,
\eeq
where the sum is over different species present in the NS: nucleons, electrons, muons and (in the presence of $\LL$ force) neutrinos.
Finally, the partial densities $\rho_i$ are related to the corresponding chemical potentials via
\beq \label{drho/dn}
\left( \frac{\d\rho_i}{\d n_i} \right)_V = \mu_i \;.
\eeq
In the absence of new forces, \Cref{hydro,P(rho),drho/dn} together with (\ref{neutr}) to (\ref{S_nb}) (with $n_{\anm}=0$) allow one to solve for the NS pressure and density profiles, determine its mass-radius relation and find the equilibrium densities of species. 
The procedure suitable for numerical implementation
is presented in Appendix~\ref{app:B}.
We will now discuss the remaining ingredients necessary to solve for the NS structure in the presence of $\LL$ force.

\subsubsection{Effect of the $\LL$ force}

To find the effect of the new force on chemical equilibrium, one needs to calculate the $\LL$ electric potential $\phi$ generated by muons, muon antineutrinos and tau neutrinos produced by the Schwinger mechanism.
Assume a static spherically-symmetric distribution of charges.
In the metric (\ref{metric}) the equation of motion for the 0-th component of the gauge boson reads as follows,
\beq \label{EqPhi2}
\begin{split}
\e^{-2\lambda-\Phi}\left( \frac{\diff^2}{\diff r^2} +  \left(  \frac{2}{r} - \Phi'  - \lambda' \right) \right. &\left. \frac{\diff}{\diff r} \right)A_0 - m_A^2A_0 \\
& = g(n_\mu-n_{\anm}-n_{\nt})  \;,
\end{split}
\eeq
where prime means derivative with respect to $r$ and we work in the Lorenz gauge $\partial_\mu A^\mu = 0$.
The differential operator in the l.h.s. of this equation is the covariant Laplacian in the metric (\ref{metric}). 
One needs to find a regular solution to this equation with the vanishing boundary condition at infinity. The $\LL$ electric potential is then determined by 
\beq\label{phiViaA}
\phi = A_0\e^{-\Phi} \;,
\eeq
see Appendix~\ref{app:A} for details.
The electric potential contributes to the chemical potentials of species charged under $\LL$:
\beq \label{MuMuP}
\mu^{(p)}(r) = \mp g\phi(r) \;,
\eeq
where minus (plus) stands for particles of equal (opposite) charge. 
Thus, the total chemical potentials of muons and muon antineutrinos read as follows,
\beq \label{MuMu12}
 \mu_{\mu} =  \mu_{\mu}^{(k)} + \mu_{\mu}^{(p)} \;, ~~~ \mu_{\anm} = \mu_{\anm}^{(k)} + \mu_{\anm}^{(p)} \;,
\eeq
where $\mu^{(k)}_{\mu,\anm}$ are given in \Cref{Mu_mu}. 
Crucially, the new long-range force prevents us from solving for chemical equilibrium locally: by \Cref{EqPhi2}, the chemical potentials of charged species at a given point depend on the global distribution of charges.

Finally, we notice that in equilibrium, where particle numbers do not change, the Schwinger pair production has stopped, meaning that $\anm$ and $\nt$ occupy all available bounded energy levels in the background potential. Hence they have vanishing Fermi energy and its (position-dependent) Fermi momentum is $|p_F|\approx g|\phi|$.
Thus, we can write
\beq\label{nAnm} 
n_{\anm}=n_{\nt}=g^3|\phi|^3 /(3\pi^2)\;
\eeq
and
\beq\label{MuAnm}
\mu_{\anm}^{(k)}=(3\pi^2 n_{\anm})^{1/3} \;.
\eeq

\begin{figure}[t]
\center{
		\begin{minipage}[h]{0.99\linewidth}
			\center{\includegraphics[width=0.99\linewidth]{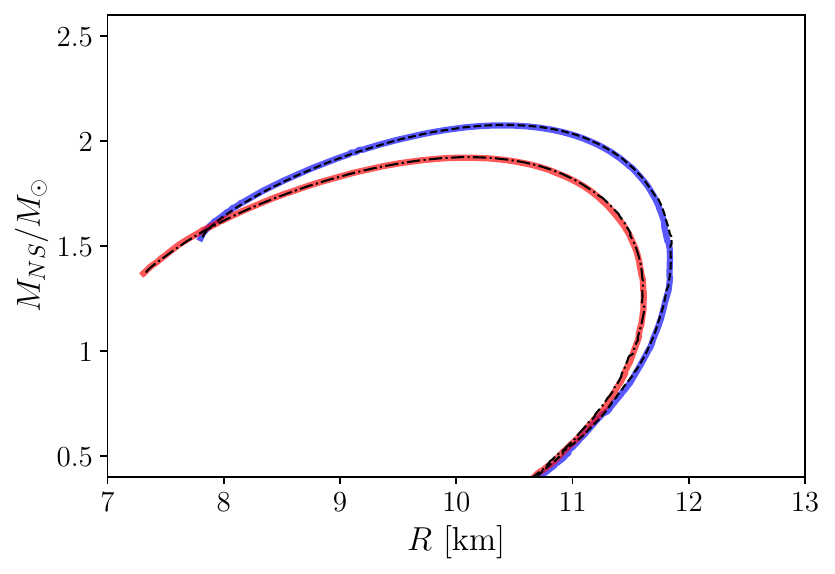}}
		\end{minipage}
	}
  \caption{The Mass-Radius relation in the presence of $\LL$ force. We vary the nuclear equation of state parameter $\gamma$ and the gauge coupling $g$. We take $(\gamma,g)=(0.13,0)$ (black dashed), $(0.13,10^{-17})$ (blue solid), $(0.2,0)$ (black dash-dotted), $(0.2,10^{-17})$ (red solid). We see no noticeable difference in the M-R curve due to the new force. The gauge boson mass is $m_A=10^{-14}$ eV in this benchmark.
  }
	\label{fig:M_R_relation}
\end{figure}

\begin{figure*}[t]
    \center{
        \begin{minipage}[h]{0.48\linewidth}
            \center{\includegraphics[width=0.99\linewidth]{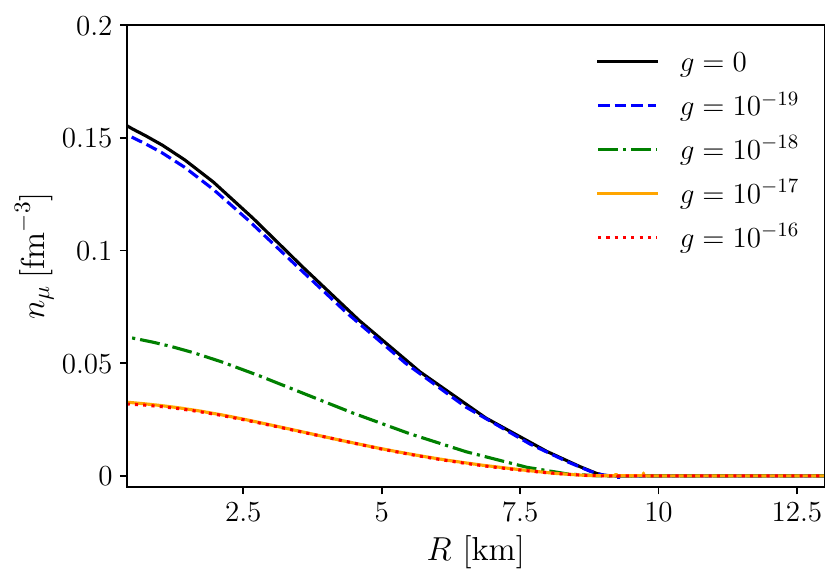}}
        \end{minipage}
		\hfill
		\begin{minipage}[h]{0.48\linewidth}
			\center{\includegraphics[width=0.99\linewidth]{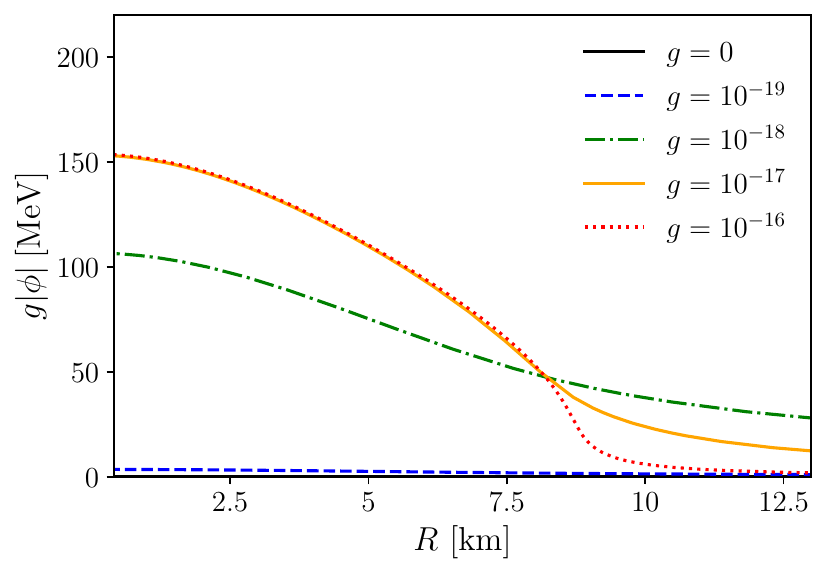}}
		\end{minipage} \\
        \begin{minipage}[h]{0.48\linewidth}
            \center{\includegraphics[width=0.99\linewidth]{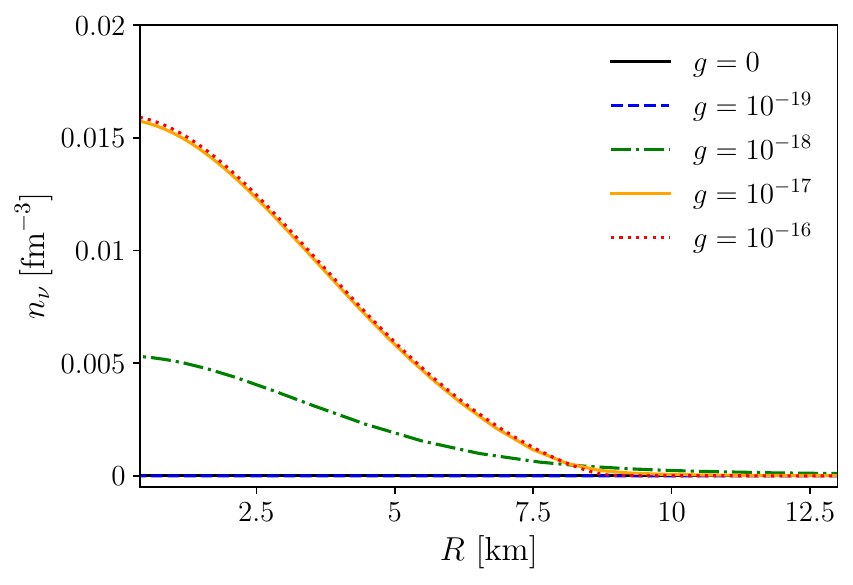}}
        \end{minipage}
		\hfill
		\begin{minipage}[h]{0.48\linewidth}
			\center{\includegraphics[width=0.99\linewidth]{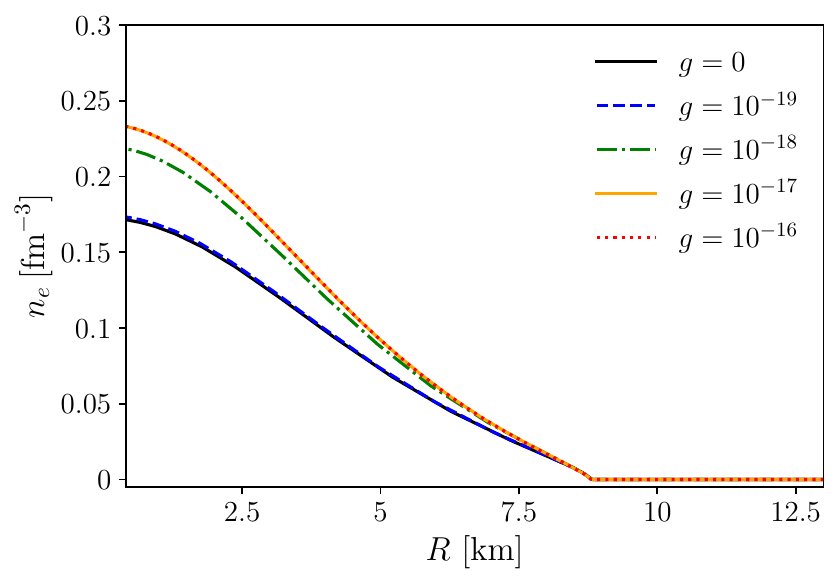}}
		\end{minipage}        
	}
\caption{Distribution of muons $n_\mu$, electrons $n_e$, neutrinos $n_{\anm}=n_{\nt}$ and the $\LL$ electric potential energy $g|\phi|$ of the NS with $M_{NS}=2M_{\odot}$ at equilibrium. We take $m_A=10^{-14}$ eV, $\g=0.13$. }
\label{fig:Dens}
\end{figure*}

We now have all ingredients necessary to calculate the effect of the new long-range force on the NS parameters.
The numerical algorithm for doing so is described in Appendix~\ref{app:B}.
Here we present results. 
We take the gauge boson mass $m_A=10^{-14}$ eV as a benchmark value, since at this (and lower) mass we have the stringent constraint on the gauge coupling from the NS binary mergers \cite{Dror:2019uea}, and, at the same time, the superradiance constraints do not apply~\cite{Baryakhtar:2017ngi}. Next, to test the robustness of our results against the variation of the nuclear equation of state, we consider two values of the parameter $\g$ controlling the stiffness of nuclear matter, $\g=0.13$ and $0.2$; see Appendix~\ref{app:B} for the definition of $\g$. Finally, the gauge coupling is varied in the range $0<g\leqslant 10^{-16}$, where the upper limit is motivated by the $\Delta N_{\rm eff}$ bound during the nucleosynthesis due to enhanced neutrino annihilation \cite{Dror:2020fbh}. Recall that, according to the estimates of Sec.~\ref{sec:neutr}, we expect the significant back-reaction of the $\LL$ force on the NS structure at $g\gtrsim 10^{-18}$. 

First, we compute the Mass-Radius relation of the NS to see, in particular, if the maximum NS mass is affected by the new force. 
The relation is shown in Fig.~\ref{fig:M_R_relation} for $g=0,10^{-17}$ and $\g=0.13,0.2$. We observe no effect due to the new force. This is expected since mass and radius are mostly determined by the distributions of protons and neutrons, and the latter are little affected by the $\LL$ force.

Next, we calculate the equilibrium distributions of $\e$, $\mu$ and $\nu$ as functions of the distance from the NS center. They are shown in Fig.~\ref{fig:Dens} for several values of $g$ in the range $0\leqslant g\leqslant 10^{-16}$.
There we also plot the $\LL$ electric potential energy $g|\phi|$ at the same values of $g$.
We see that $g|\phi|$ reaches $\sim 100$ MeV in the NS interior at $g\sim 10^{-18}$, and that at this value it starts affecting the distributions of leptonic species, which agrees with the estimate made in Sec.~\ref{sec:neutr}.
We observe a substantial decrease of the number density of muons and an increase of the electronic number density.
The reason is clear: since muons are repelled in the $\LL$ potential, less of them are present in equilibrium and, correspondingly, more electrons remain unconverted by the weak reactions.
As a result, the total $\LL$ charge of the NS is significantly reduced compared to the case without the new force.
Besides, $\anm$ and $\nt$ form degenerate Fermi-gas around the NS, additionally screening the charge. 

\begin{figure}[h]
\center{
		\begin{minipage}[h]{0.99\linewidth}
			\center{\includegraphics[width=0.99\linewidth]{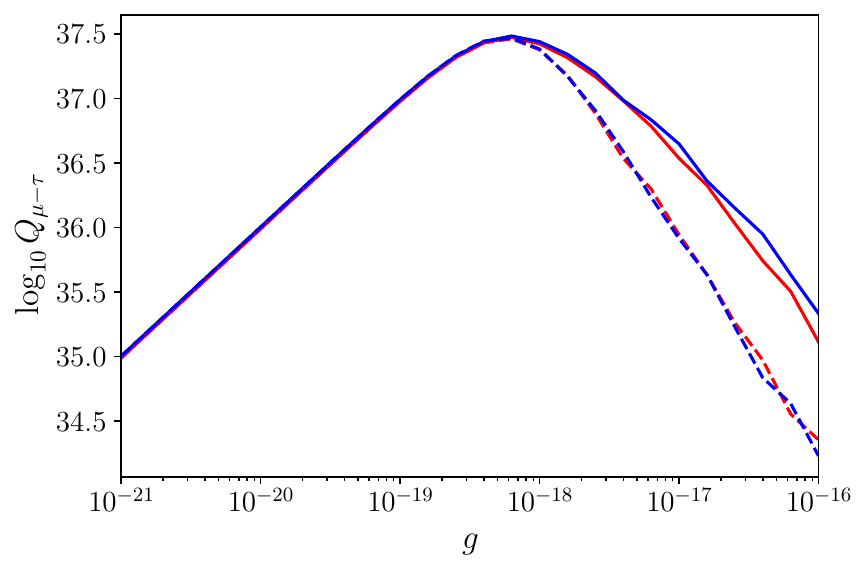}}
		\end{minipage}
	}
  \caption{The total $\LL$ charge of the NS of mass $2M_{\odot}$ measured at the distance 10 km (solid lines) and 30 km (dashed lines) from its center, as a function of the $\LL$ coupling $g$. The blue lines correspond to the stiffness parameter $\g=0.13$ and the red lines to $\g=0.2$. We take $m_A=10^{-14}$ eV. The wiggles in the lines are due to the limiting accuracy of the numerical algorithm used to find the NS equation of state; see Appendix~\ref{app:B} for details.}
	\label{fig:screening}
\end{figure}

The screening effect is demonstrated in Fig.~\ref{fig:screening}, which shows the $\LL$ charge $Q=gN$, where $N$ is the net number of charged particles, measured at the surface of the NS with $R_{NS}\approx 10$ km, at different values of $g$. 
To see how the screening effect behaves as a function of distance, we also plot the charge measured at the distance of $3R_{NS}$ from the NS center.
We see that $Q\propto g$ at $g\ll 10^{-18}$. Meanwhile, with  larger couplings, the back-reaction kicks in and changes $N$.

\subsection{Approach to equilibrium}
\label{ssec:sn}

We showed above that the equilibrium state of the NS in the presence of $\LL$ force is markedly different from the state without the new force as soon as $g\gtrsim 10^{-18}$. In particular, it features the ``Fermi-balls'' of $\anm$ and $\nt$ screening partially the $\LL$ charge. 
Let us discuss how this final state can be achieved during the evolution of the star following the progenitor supernova explosion. 
Solving exactly for the chemical potentials and element abundances in the out-of-equilibrium environment is a difficult problem, and our discussion here is only qualitative.

During the NS evolution its chemical composition changes following the direction of out-of-equilibrium reactions (\ref{n_pe}), (\ref{n_pmu}).
The timeline begins at $t=0$ corresponding to the core bounce; the relevant species are indicated below in parentheses.

\textit{(i) --- ($p, n, e$) --- } We start at the point when muons are not in weak equilibrium, and their fraction is well below the electron one (see e.g. \cite{Fischer:2020vie}). The condition (\ref{Mu_npemu}) is not satisfied since $\mu_n-\mu_p \gtrsim \mu_e > \mu_\mu$.
The electron chemical potential increases until it reaches the muon mass $m_\mu\simeq 106$ MeV.

\textit{(ii) --- ($p, n, e, \mu$) --- } At the second stage the yield of muons increases due to the out-of-equilibrium processes (\ref{n_pmu}), while muon anti-neutrinos mostly escape the star with the mean energy $E_\nu\sim 10$ MeV~\cite{haensel2007neutron}.
As a result, the star acquires the $\LL$ charge.
The charge generates new contributions $\mu_{\mu}^{(p)} = - \mu_{\anm}^{(p)}$ to the chemical potentials of $\mu$ and $\anm$, see \Cref{MuMuP,MuMu12}. 
Both electron and muon chemical potentials continue to increase until the moment when the $\LL$ electric potential is strong enough to trap a significant fraction of muon antineutrinos.
This happens before the beta-equilibrium could be established, since
$\mu_n-\mu_p\gtrsim \mu_\mu \gtrsim m_\mu > E_\nu$.

\textit{(iii) --- ($p, n, e, \mu, \anm$) --- } The third stage is characterized by the enhanced, compared to the case without the $\LL$ force, population of $\anm$ in the star.
Newly produced muons and most of the muon antineutrinos remain in the star. Some of $\anm$ with higher-than-average energy nevertheless escape, and the $\LL$ charge continues to grow.
Later, new muonic force would play a significant role in the evolution of lepton abundances, which continues until the beta-equilibrium \Cref{Mu_npemu} is established. Compared to the case without the new force, the beta-equilibrium features the reduced number of muons and a population of muon antineutrinos trapped in the $\LL$ potential well. The depth of the potential at this stage is determined by the difference $n_\mu-n_{\anm}$; see \Cref{EqPhi2}.

\textit{(iv) --- ($p, n, e, \mu, \anm, \nt$) --- } During the final stage, which is much longer compared to the previous stages (see \Cref{tQ}), the Schwinger mechanism supplies the NS with more muon antineutrinos and tau neutrinos. Thus, the $\LL$ charge of the NS keeps diminishing. The neutrino pair production stops when all available levels are occupied. The final number densities of $\anm$ and $\nt$ in the potential $\phi$ are given by \Cref{nAnm}; note that $\phi$ itself must be found consistently in the presence of $\mu$, $\anm$ and $\nt$, as follows from \Cref{EqPhi2}. The equilibrium is finally reached. 

\begin{figure}[b]
    \centering    \includegraphics[width=1\linewidth]{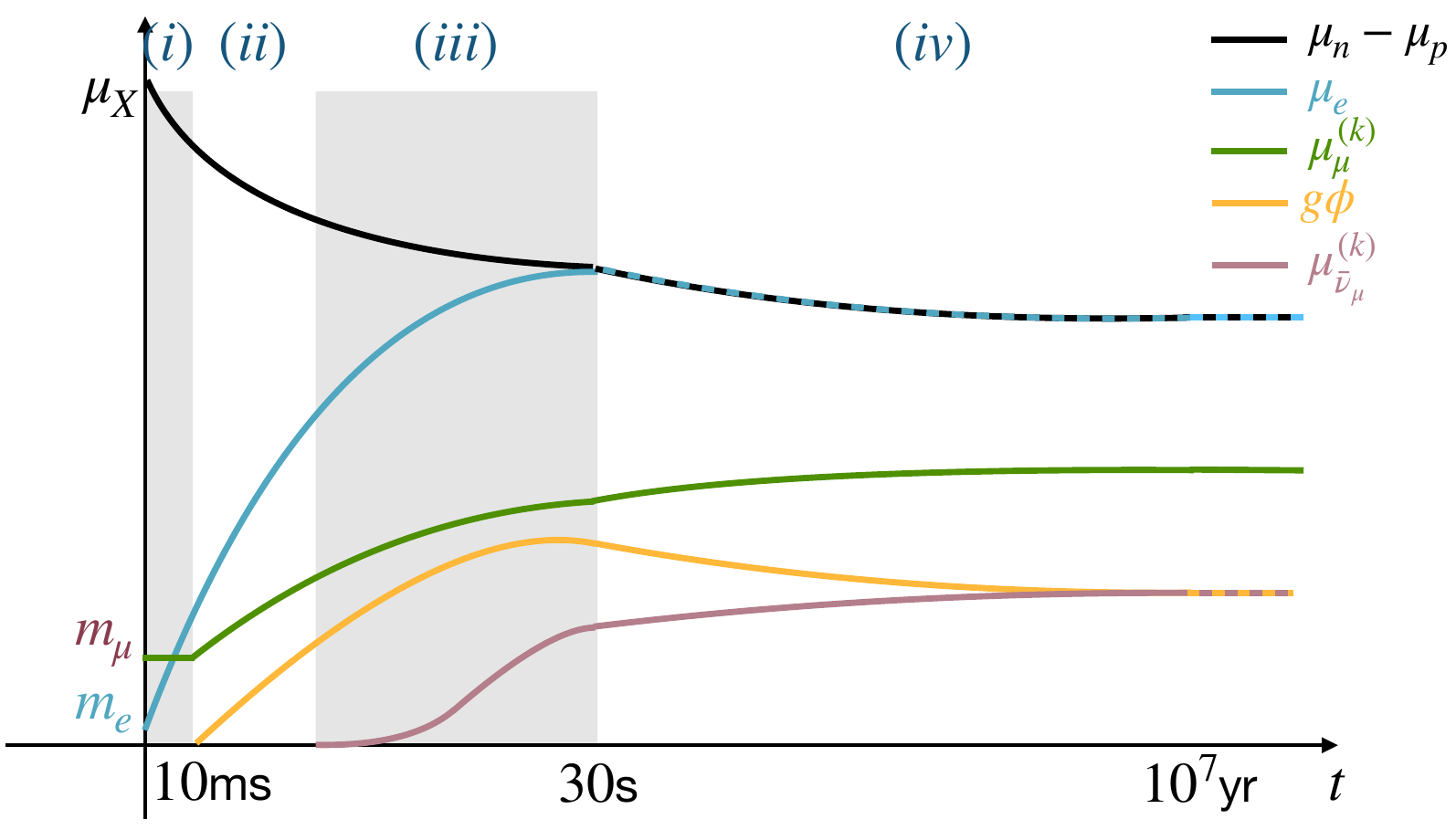}
    \caption{Schematic plot showing the evolution of various chemical potentials, from the NS birth until the equilibrium is reached, in the presence of the $\LL$ force with strong back-reaction on the NS leptonic abundances ($g\gtrsim10^{-18}$). The change of the $\LL$ electric potential energy $g\phi$ is also shown.
See the text for the explanation of the stages \textit{(i)--(iv)}. 
}
    \label{fig:evol}
\end{figure}

In Fig.~\ref{fig:evol} we plot schematically the evolution of the chemical potentials $\mu_n-\mu_p$, $\mu_e$, $\mu_\mu$, $\mu_{\anm}$ and of the potential energy $g|\phi|$ at the stages \textit{(i)--(iv)}.

An important comment is in order. The estimate (\ref{tQ}) for the duration of stage \textit{(iv)}, i.e. the time of discharge, follows from the naive estimate of the depth of the potential well $|\phi_{\rm max}|\propto g$. This estimate is applicable to a NS at $g\lesssim 10^{-18}$, when the back-reaction of the gauge force on the NS equation of state is small. As Fig.~\ref{fig:Dens} demonstrates, once the back-reaction kicks in, the depth of the potential well is quickly saturated, so that $g|\phi_{\rm max}|\sim 100\, {\rm MeV}$ at all $g\gtrsim 10^{-18}$. 
This means that the $\LL$ electric field and the Schwinger neutrino flux are also saturated, hence one can estimate $t_Q\sim 10^7$ yr. for $g\gtrsim 10^{-18}$.

\section{Phenomenological consequences}
\label{sec:pheno}

\subsection{Bound on $g$ from NS mergers}
\label{ssec:binaries}

\begin{figure}
    \centering    \includegraphics[width=0.9\linewidth]{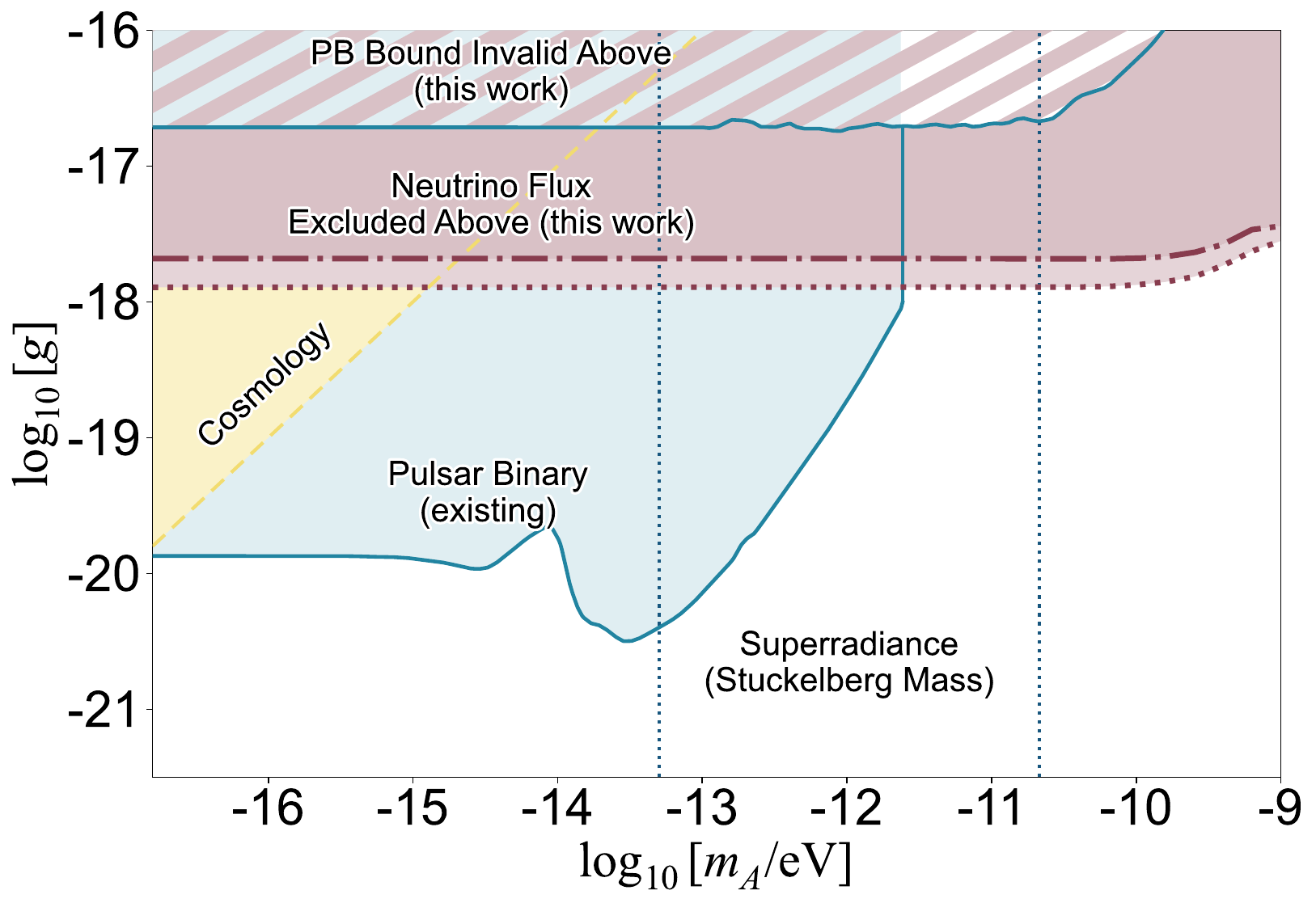}
    \caption{The parameter space for $\LL$ gauge coupling $g$ and gauge boson mass $m_A$. The hatched region indicates that the suppression of the NS muonic charge due to the back-reaction of the $\LL$ force considered here is so strong, that the pulsar binary constraint~\cite{Dror:2019uea} (blue) should be relaxed. The cosmological constraint (yellow) is dominated by enhanced neutrino annihilation~\cite{Dror:2020fbh}. Superradiant constraint from stellar-mass black holes is shown in between dotted blue lines~\cite{Baryakhtar:2017ngi}, while it should only apply to non-Higgsed gauge boson mass~\cite{East:2022rsi}. 
    Region above the red lines is prospectively excluded by non-observation of the Schwinger neutrino flux from a young nearby NS. We assume a targeted neutrino telescope aiming at the young isolated star RX J1856.5-3754 located at the distance $0.123^{+0.011}_{-0.015}\,$kpc~\cite{walter2010revisiting}. The dash-dotted (dotted) red line corresponds to the maximum (minimum) distance to the NS.
    }
    \label{fig:constraint1}
\end{figure}

Ref.~\cite{Dror:2019uea} establishes stringent constraints on the $\LL$ gauge coupling by studying how the dipole radiation of the gauge boson during the NS-NS and NS-BH mergers modifies the gravitational-wave signal. For an optimistic assumption about the muonic fraction in the NS, Ref.~\cite{Dror:2019uea} quotes $g\gtrsim 3\cdot 10^{-21}$ (NS--NS) and $g\gtrsim 10^{-21}$ (NS--BH) at $m_A\lesssim 10^{-14}$ eV. Our analysis shows that these merger constraints do not apply for $g\gtrsim 10^{-17}$ because of the reduced number of muons in equilibrium and charge screening by the neutrino Fermi-balls. This opens a region $10^{-17}\lesssim g$ as shown in Fig.~\ref{fig:constraint1}.\footnote{We could reliably calculate the modified NS equation of state for $m_A$ up to $10^{-9}$ eV, which explains the upper limit for $m_A$ in Fig.~\ref{fig:constraint1}; see Appendix~\ref{sec:Numerical method}.} The upper limit now comes from the enhanced neutrino annihilation to light gauge bosons during nucleosynthesis \cite{Dror:2020fbh}.\footnote{This bound assumes that the universe was reheated to temperatures above the muon mass.}

\subsection{Bound on $g$ from a nearby supernova neutrino afterglow}
\label{ssec:flux}

Having just opened the window of the gauge couplings unconstrained by the NS binary mergers, we will now discuss how it can be prospectively closed again
from non-observation of the neutrino flux emitted by nearby NSs undergoing the process of charge neutralisation (stage \textit{(iv)} in Sec.~\ref{ssec:sn}). 
Indeed, the estimate (\ref{NuFlux}) shows that a single NS $\simeq 100$ pc away generates a flux of neutrinos of typical energy $E_\nu\sim 100$ MeV that is comparable or exceeds the observed atmospheric flux in the direction of the NS if $g\gtrsim 10^{-18}$. 
The excess flux can be detected if the angular resolution of a neutrino detector is at least several tens of degrees, as we will see momentarily.
Here we evaluate the flux more carefully, using the calculated $\LL$ electric field profile of the NS.

We make two important simplifying assumptions. First, we use the $\LL$ potential in equilibrium, i.e., when the Schwinger production stops. As discussed in Sec.~\ref{ssec:sn}, calculating the out-of-equilibrium potential is a challenging task. Using the equilibrium potential is a good conservative approximation at times $t\sim t_Q$ when, on the one hand, the NS is close to equilibrium and, on the other hand, the Pauli blocking does not yet suppress the Schwinger production. 
Second, we neglect the general relativistic effects on the neutrino propagation. As discussed in Appendix~\ref{app:A}, including these effects will correct our results by $\sim 10\%$. 

The typical momentum gained by the neutrino ($\nm$ and $\ant$) during its acceleration along the gradient of the $\LL$ potential is $p_r\sim 100$ MeV. On the other hand, the momentum spread transverse to the gradient of the potential 
\cite{Cohen:2008wz} is $\Delta p = \sqrt{g\E/(2\pi)}\lesssim \sqrt{-g\phi/R_{NS}}\sim 10^{-7}\:\MeV \ll p_r$.
Thus, we only need to consider particles moving radially from the NS center. 

Under these simplifications, the neutrino flux (``afterglow'') is given by \Cref{NuFlux}, where 
\beq\label{dNdt2}
\frac{\diff N_t}{\diff t}=2\cdot 4\pi\limitint_0^{\infty}\diff r\:r^2\Gamma(r) \;,
\eeq
and the Schwinger rate is given in \Cref{SchwRate} (without the exponential suppression factor, according to \Cref{ma1,ma2}), with $\E(r)=-\phi'(r)$ and $\phi(r)$ calculated in Sec.~\ref{ssec:eos}. An additional factor 2 in \Cref{dNdt2} counts two produced neutrino flavors.
To find the differential flux, we assumed no momentum change for the neutrinos, except for the radial acceleration due to the $\LL$ potential. So if a neutrino is produced with zero initial momentum at radius $r$, 
its final momentum at Earth (infinity) is
$ E_\nu(r) = -g\phi(r)$. Thus, the differential flux at Earth is given by
\beq
\frac{\diff N_t}{ \diff t\, \diff E_\nu\, \diff A} = \frac{2}{4\pi L^2}\times \left.4\pi r^2 \Gamma(r)\frac{1}{g|\phi'(r)|}\right|_{g|\phi(r)| = E_\nu} \;.
\eeq
Its flavour composition can be estimated as 
$\nu_e:\nu_\mu:\nu_\tau \sim 1:2:2$ for the original muon neutrino and $\ane:\anm:\ant \sim 1:1.5:1.5$ for the original tau antineutrino \cite{ParticleDataGroup:2024cfk}.\footnote{The $\LL$ potential of the NS freezes the flavours of Schwinger-produced neutrinos, which propagate adiabatically while inside the potential and experience vacuum oscillations outside it.}

\begin{figure}[t]
    \centering
    \includegraphics[width=8cm]{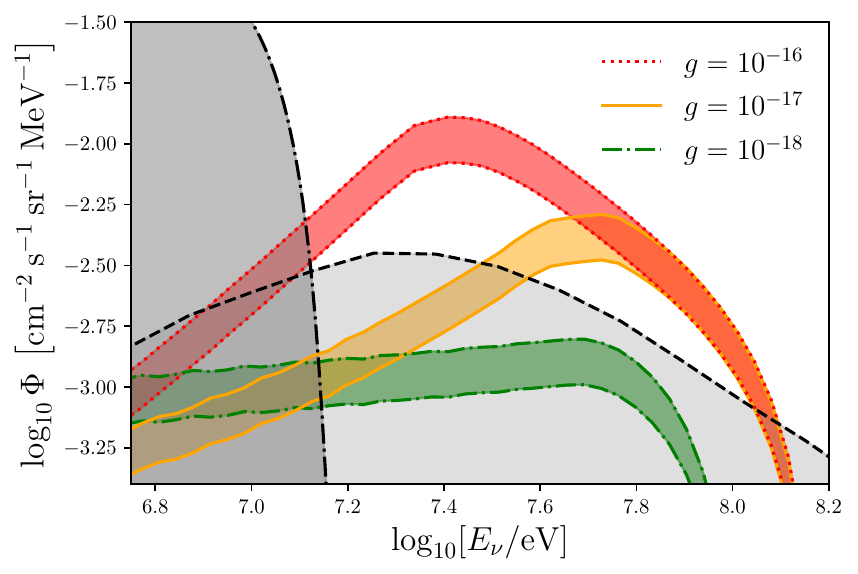}
    \caption{Differential neutrino flux generated by the Schwinger mechanism 
    from the neutron star RX J1856.5-3754, smeared over the area with $\theta_{\text{res}}= 23^\circ$~\cite{Alhazmi:2016qcs,Suzuki:2019jby,Bodur:2021nueO16Indico},  
    at several values of the $\LL$ gauge coupling $g$.
    The black dashed and dash-dotted lines denote the atmospheric and solar neutrino spectra, respectively~\cite{Gaisser:2002jj,Vitagliano:2019yzm}.
    The gauge boson mass is $m_A=10^{-13}$ eV.
    The bands show the systematic uncertainties, the main of which is the uncertainty in the distance to the NS.
    }
    \label{fig:Neutrino_Flux}
\end{figure}

As a concrete example of a young nearby NS, we take the young isolated star RX J1856.5-3754, located 
$0.123^{+0.011}_{-0.015}\,$kpc away \cite{walter2010revisiting}.
Its age is estimated as $\lesssim 5\cdot 10^5$~yr.~\cite{mignani2013birthplace}, hence in the presence of the $\LL$ force it would be now at the charge neutralization stage.
Fig.~\ref{fig:Neutrino_Flux} shows the expected differential neutrino flux from this NS at several values of $g$ and the gauge boson mass $m_A=10^{-13}\,\text{eV}\ll R_{NS}^{-1}$.
We adopt $\theta_{\text{res}}= 23^\circ$ as a benchmark angular resolution of the neutrino detector such as Super-Kamiokande at $E_\nu\sim100$ MeV~\cite{Alhazmi:2016qcs,Suzuki:2019jby,Bodur:2021nueO16Indico}.
The uncertainty bands correspond to the uncertainty in the distance to the NS; other sources of systematic uncertainties are expected to be at the level of $\sim 10\%$.

We observe that at $g=10^{-18}$ the signal is comparable to the atmospheric neutrino flux at $E_\nu\sim 100$ MeV, and at higher values of $g$ the signal exceeds the atmospheric flux. This allows us to place the prospective bound on Fig.~\ref{fig:constraint1} that would rule out the region $g\gtrsim 10^{-18}$ at $m_A\lesssim 10^{-10}\:\eV\sim R_{NS}^{-1}$. At higher values of $m_A$, the $\LL$ potential and the production rate are suppressed by powers of $m_AR_{\rm NS}$ (see Appendix~\ref{app:A}), and the bound relaxes.
A dedicated search for the neutrino afterglow from RX J1856.5-3754 is needed to firmly establish the bound. 

\section{Conclusion}
\label{sec:concl}

In this paper we studied possible effects of new long-range forces resulting from their interactions with the active neutrino species. We considered the forces arising from gauging the SM global symmetries: $B-L$, $L_e-L_{\mu/\tau}$ and $\LL$. Astronomical bodies, such as planets or neutron stars, source large electric fields associated with these forces. These fields can produce active neutrino pairs by the Schwinger mechanism. We estimated that for the $B-L$ and $L_e-L_{\mu/\tau}$ forces this effect is too weak to be observable or to challenge the existing bounds on the gauge couplings from fifth-force experiments. On the other hand, the $\LL$ electric field around NSs can be strong enough to generate an observable neutrino flux.

Furthermore, we showed that the $\LL$ force can impact the NS equation of state, as anticipated in Ref.~\cite{Dror:2019uea}. We calculated the equilibrium abundances of leptonic species in the presence of the $\LL$ electric potential, at various values of the gauge coupling $g$. The back-reaction of the new force becomes significant at $g\gtrsim 10^{-18}$. A notable effect is the reduced number of muons in equilibrium, leading to the reduced $\LL$ charge. The latter is additionally screened by muon antineutrinos and tau neutrinos accumulated around the NS. This makes the bound on $g$ from the dipole radiation in NS-NS and NS-BH binary mergers~\cite{Dror:2019uea} inapplicable at $g\gtrsim 10^{-17}$.

Neutrinos provide an indispensable probe of the inner region of proto-neutron stars on their way from a progenitor supernova to the NS~\cite{Li:2020ujl}.
These objects are, however, transient and scarce.
We showed that in the presence of the new force, NSs may have a neutrino ``afterglow'' -- a steady neutrino flux generated by the Schwinger mechanism and unfading on a time scale $t_Q\sim 10^7$ yr (see \Cref{tQ}), long after the NS formation.
A single young NS located $\sim 100$ pc away would generate an observable excess of neutrino with the typical energy $E_\nu\sim 100$ MeV, at $g\gtrsim 10^{-18}$. 
As a concrete example, we took one of the ``magnificent seven'' star RX J1856.5-3754, known to be much younger than the discharge time $t_Q$, and plotted the expected signal on Fig.~\ref{fig:Neutrino_Flux}. This suggests that one can look for the new gauge force with a dedicated directional neutrino signal search from nearby NSs.

It is important to understand how the equilibrium state of the NS, modified by the new gauge force, is actually reached during the NS evolution beginning with the progenitor supernova explosion. In this paper, we only sketched the main stages of the evolution; a more careful analysis would be phenomenologically interesting.
For example,
muons can play an important role in the NS cooling at late times ($t>10$ s after the core bounce) \cite{Sugiura:2022cwi}.
This stage of the NS evolution might be accessible to observations for the next Galactic supernova \cite{Li:2020ujl}.
Investigating the impact of the new gauge force on the NS cooling is an interesting direction of research~\cite{Manzari:2023gkt,Fiorillo:2026wso,Fiorillo:2026dqu}.

Another example of the NS dynamics, which may be relevant for the search for new gauge forces, is binary stars in which the NS accretes matter from a companion star. The accretion leads to the NS being displaced from beta-equilibrium~\cite{Haensel:2007wy}. This can potentially re-ignite the Schwinger neutrino production in old NSs, leading to new opportunities in detecting the neutrino ``afterglow''.

Finally, it would be interesting to study further phenomenological implications of the neutrino pair production by strong fields in theories beyond the SM and in various environments~\cite{Tanin:2026crp}.

\textbf{Note added.} After completion of this work, Refs.~\cite{Fiorillo:2026wso,Fiorillo:2026dqu} appeared that
put new bounds on the muonic force from NS cooling. 
Assuming that the muonic force does not change the NS equation of state, 
they find that the NS cooling provides the strongest bound on the gauge coupling in the gauge boson mass range $10^{-5}$ eV $\lesssim m_A\lesssim 10^5$ eV, which is complementary to the mass range studied in this work.

\section*{Acknowledgments}

We thank Marco Costa, Jeff Dror, Junwu Huang, Maxim Pospelov and Sergey Sibiryakov for useful discussions.
Research at Perimeter Institute is supported in part by the Government of Canada through the Department of Innovation, Science and Economic Development Canada and by the Province of Ontario through the Ministry of Colleges and Universities. 
This work was supported by the National Natural Science Foundation of China No.~12450006., the Department of Energy under Grant No.~DE-SC0011842 and in part by a Sloan Research Fellowship from the Alfred P. Sloan Foundation at the University of Minnesota.

\appendix

\section{}
\label{app:A}
\subsection{Electric potential of a homogeneously-charged body}

Consider a spherically symmetric uniformly charged body with radius $R$ and charge density $gn$.
In flat space, the electric potential $\phi\equiv A^0$ satisfies the equation
\beq 
\phi''+\frac{2}{r}\phi'-m_A^2\phi = -gn \;.
\eeq
The solution vanishing at infinity reads as follows
\beq 
\begin{aligned}
& \phi(r)=gn\varphi(r) \;, \\
& \varphi(r) = \left\lbrace \begin{array}{ll}
    \frac{m_Ar-(m_AR+1)\sinh(m_Ar)\e^{-m_AR}}{m_A^3r}  \;, & r< R  \\
    \frac{(m_AR\cosh(m_AR)-\sinh(m_AR))\e^{-m_Ar}}{m_A^3r} \;,  & r>R  
\end{array} \right.
\end{aligned}
\eeq
In the limit $m_AR\ll 1$ the potential around the body reduces to the Coulomb potential.
The electric field strength $\E_r(r)=-\phi'(r)$ reaches maximum at the surface of the body,
\beq 
|\E_{max}^{(m_AR\ll 1)}| = \frac{gnR}{3} \;.
\eeq
In the opposite limit, $m_AR\gg 1$, the potential inside the body tends to constant, $\phi(r<R)\approx gn/m_A^2$, and outside the body it vanishes.
At $r\sim R$, the electric field strength is given by
\beq 
\E_r = \frac{gnR}{2m_Ar}\e^{-m_A|r-R|} \;,
\eeq
up to exponentially small corrections.
The maximum of $\E_r$ is attained at $r\approx R$ and equals
\beq
|\E_{max}^{(m_AR\gg 1)}|\approx \frac{3|\E_{max}^{(m_AR\ll 1)}|}{2m_AR} \;,
\eeq
which is suppressed compared to the Coulomb limit.

\subsection{Electric potential of a neutron star}

We start from the covariant generalization of the Lagrangians (\ref{L1}), (\ref{L2}).
The Lagrangian of the vector field reads as
\beq 
\L_A = \sqrt{-\text{g}}\left( -\frac{1}{4}F_{\mu\nu}F^{\mu\nu} +\frac{1}{2}m_A^2A_\mu A^\mu \right) \;,
\eeq
where the indices are raised and lowered with the metric $\text{g}_{\mu\nu}$.
The interaction term is written as
\beq 
\L_{int} = -\sqrt{-\text{g}}\, \text{g}^{\mu\nu}e_{a\nu}gA_\mu \bar{\psi}\gamma^a\psi \;,
\eeq
where $e_{a\nu}$ is the tetrad field and for simplicity we write the coupling to a single fermionic species.
We focus on the $0$'th component of the fermionic current identified with the number density of charge carriers, $n=\bar{\psi}\gamma^0\psi$.
We substitute the metric ansatz (\ref{metric}) to the action $S=\int\diff^4x(\L_A+\L_{int})$ and vary the action with respect to $A_0$.
This leads to \Cref{EqPhi2}.

In practice, the equation for $A_0$ can be solved iteratively.
In the flat space limit, the solution with the vanishing boundary condition at infinity reads
\beq 
\begin{split}
A_0^{(0)}(r) &  = - g\limitint_0^r\diff r'\frac{r'n(r')\sinh m_Ar'\cdot \e^{-m_Ar}}{m_Ar} \\
& - g\limitint_r^{\infty}\diff r'\frac{r'n(r')\sinh m_Ar\cdot \e^{-m_Ar'}}{m_Ar} \;.\label{staticElectric}
\end{split}
\eeq
The first correction $A_0^{(1)}$ due to the curved metric still satisfies the flat space limit of \Cref{EqPhi2} but with the r.h.s. replaced by $-\Delta_g^{(1)}A_0^{(0)}$ where 
\beq 
\Delta_g^{(1)}=-(2\lambda+\Phi)\left( \frac{\diff^2}{\diff r^2} + \frac{2}{r}\frac{\diff }{\diff r} \right) - (\lambda'+\Phi')\frac{\diff }{\diff r} \;.
\eeq

In curved spacetime the electric potential $\phi$ is extracted from the decomposition of the vector field into longitudinal and transverse components,
\beq
A_\mu = \phi u_\mu + a_\mu ~~\Rightarrow~~ \phi = - A_\mu u^\mu \;,
\eeq
where $u_\mu$ is the field of static observers, $u^\mu u_\mu = -1$, and $a^\mu u_\mu =0$.
In our case $u^\mu = (u^0,\vect{0})$ with $u^0=-\e^{-\Phi}$ and $A_\mu=(A_0,\vect{0})$, hence
\beq\label{PhiA0}
\phi = A_0\e^{-\Phi} \;.
\eeq
Next, the electric $\E_\mu$ and magnetic $\mathcal{B}_\mu$ field strengths are read off from the decomposition of the field tensor,
\beq 
F_{\mu\nu} = 2 u_{[\mu}\E_{\nu]}+\epsilon_{\mu\nu\sigma}\mathcal{B}^{\sigma} \;.
\eeq
Projecting onto the first term and using \Cref{PhiA0}, we obtain 
\beq
\E_\mu = -h_\mu^\nu\d_\nu\phi ~~ \Rightarrow ~~ \E_r = -\phi' \;,
\eeq
where $h_\mu^\nu$ is the spatial metric tensor and prime means derivative with respect to $r$.
Finally, the electric chemical potential is defined as the work done to move a particle from infinity to a given position against the electric field,
\beq 
\mu^{(p)}(r) = \mp g\limitint_{\phi(\infty)}^{\phi(r)}\diff\phi = \mp g\phi(r) \;,
\eeq
where the minus (plus) sign stands for the attractive (repulsive) interaction.
To compute $\mu^{(p)}(r)$, one can expand $\phi(r)$ around the flat space limit $\phi^{(0)}(r)$. Then, from \Cref{PhiA0} we have
\beq 
\phi^{(0)} = A_0^{(0)} \;, ~~ \phi^{(1)}=-\Phi A_0^{(0)} + A_0^{(1)} \;.
\eeq

\begin{figure}[t]
    \centering
    \includegraphics[width=8cm]{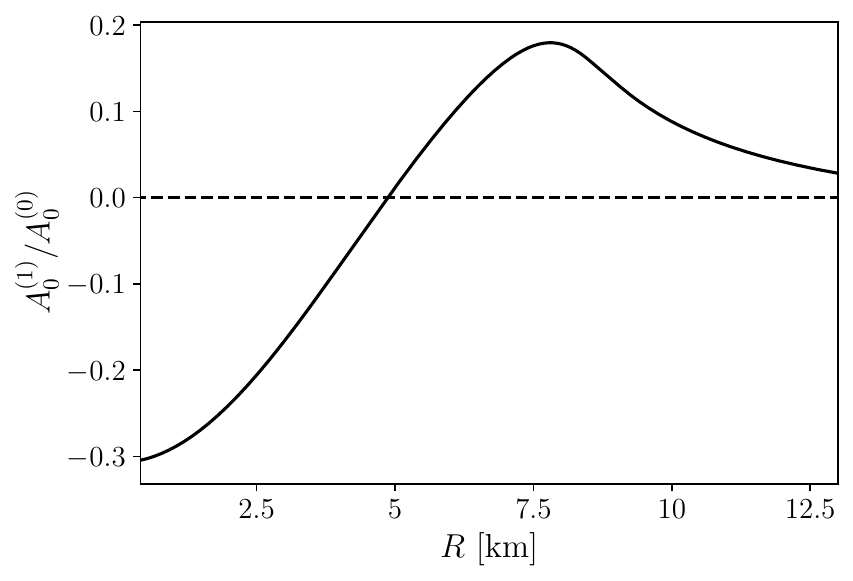}
    \caption{The GR correction to the $0$'th component of the gauge boson field inside the NS. We take $m_A=10^{-13}$ eV, $g=10^{-17}$, $\g=0.13$ and the NS mass $M=2M_{\odot}$. }
    \label{fig:GR_modification}
\end{figure}

Fig.~\ref{fig:GR_modification} shows that the correction due to the curved metric turns out to be small, and we can neglect it in numerical calculations.

\section{}
\label{app:B}
\subsection{Equation of state of a neutron star}

Here we provide more discussion of the equilibrium state of a NS and 
how the presence of the $\LL$ force modifies this state.
We start from nuclear matter and derive the relation (\ref{Mu_np}).
Given the energy density $\rho_N(n_n,n_p)$ of nucleons, one can find the chemical potentials of neutrons $n$ and protons $p$ as
\beq \label{MunMup}
\mu_n = \left( \frac{\d\rho_N}{\d n_n} \right)_V \;, ~~ \mu_p = \left( \frac{\d\rho_N}{\d n_p} \right)_V \;.
\eeq
It is convenient to switch from $n_n$, $n_p$ to the baryon number $n_b=n_p+n_n$ and the asymmetry factor $\delta=(n_n-n_p)/n_b$.
Introduce also the energy density per baryon $\tilde\rho_N = \rho_N/n_b$.
Then the nuclear equation of state can be written as \cite{haensel2007neutron}
\beq \label{RhoTilde}
\tilde{\rho}_N(n_b,\delta) = E_0(n_b/n_0) + S(n_b/n_0)\delta^2 + m_p + \frac{1+\delta}{2}\Delta m \;,
\eeq
where $\Delta m = m_n - m_p$ and $n_0=0.15\:\text{fm}^{-3}$
is the nuclear saturation density scale.
We assume that the nuclear equation of state is not affected by the new leptonic force.
The symmetric nuclear matter energy $E_0$ and symmetry energy $S$ are parametrised as follows,
\beq 
E_0(z)=E_0\frac{z(2+\gamma-z)}{1+\gamma z} \;, ~~ S(z)=S_0 z^\zeta \;.
\eeq
The parameters are $E_0 = -15.8\:\MeV \;, ~~ S_0=32\:\MeV \;, ~~ \zeta=0.6$  ~\cite{Akmal:1998cf,Haensel:2007yy}.
The parameter $\gamma$ regulates the stiffness of the equation of state. We adopt $\gamma=0.13$ as a benchmark value conforming with the results of Refs.~\cite{Akmal:1998cf,Haensel:2007yy}, but we allow it to vary between $0.13$ and $0.2$ to see if our results are robust against the change of the nuclear equation of state.
From \Cref{MunMup,RhoTilde} we obtain
\beq 
\begin{split}
\mu_n - \mu_p &  = n_b\left( \frac{\d\tilde{\rho}_N}{\d n_n} - \frac{\d\tilde{\rho}_N}{\d n_p} \right) = n_b S\bigl(\frac{n_b}{n_0}\bigr) \left( \frac{\d\delta^2}{\d n_n} - \frac{\d\delta^2}{\d n_p} \right)  \\ 
& + \frac{n_b\Delta m}{2} \left( \frac{\d\delta}{\d n_n} - \frac{\d\delta}{\d n_p} \right) = 4\delta S\bigl(\frac{n_b}{n_0}\bigr) + \Delta m \;,
\end{split}
\eeq
which is \Cref{Mu_np}.
Combining it with \Cref{Mu_e,Mu_mu}, we express all chemical potentials in terms of 4 variables $n_b$, $\delta$, $n_e$, $n_\mu$.

Chemical equilibrium (\ref{Mu_npemu}) and charge neutrality (\ref{neutr}) provide 3 equations for 4 variables.
These must be supplemented with the equations of hydrostatic equilibrium (\ref{hydro}), which involve the total energy density 
\beq \label{rhoTotal}
\rho = \rho_e(n_e) + \rho_\mu(n_e,n_b) + \rho_N(n_b,\delta) \;.
\eeq
The contribution from ultrarelativistic electrons is obtained by integrating the Fermi--Dirac distribution $f_{FD}^{(e)}(p)$ in the zero-temperature approximation:
\beq 
\rho_e(n_e) = \frac{2}{(2\pi)^3}\int \diff^3\textbf{p}\; p f_{FD}^{(e)}(p) = \frac{3^{4/3}\pi^{2/3}}{4}n_e^{4/3} \;.
\eeq
The muonic energy is split into the sum of two contributions,
\beq 
\rho_\mu(n_e,n_b) = \rho_\mu^{(k)}(n_\mu) + \rho_\mu^{(p)}(n_\mu, n_b) \;.
\eeq
The first term is the local contribution calculated similarly to the electron's one,
\begin{widetext}
\beq 
\rho_\mu^{(k)}(n_\mu)  = \frac{2}{(2\pi)^3}\int \diff^3\textbf{p} \sqrt{p^2+m_\mu^2} f_{FD}^{(\mu)}(p)  = \frac{1}{8\pi^2}\left[ p_F^{(\mu)} \left( m_\mu^2 + 2\bigl(p_F^{(\mu)}\bigr)^2\right)\sqrt{\bigl(p_F^{(\mu)}\bigr)^2+m_\mu^2} - m_\mu^4 \text{arcsinh}\bigl( \frac{p_F^{(\mu)}}{m_\mu} \bigr) \right] \;,
\eeq    
\end{widetext}
where $p_F^{(\mu)}=(3\pi^2n_\mu)^{1/3}$ is the muon Fermi momentum
and we took into account that muons are only mildly relativistic.
The second term depends on the $\LL$ electric potential $\phi$,
\beq 
\rho_\mu^{(p)}(n_\mu, n_b) = n_\mu\mu_\mu^{(p)} \;,
\eeq
where the muon ``electric'' chemical potential $\mu_\mu^{(p)}$ is given in \Cref{MuMuP}. This ``electric'' chemical potential always cancels with its counterpart of the muon antineutrinos, and its contribution effectively contributes to Eq.~(\ref{Mu_npemu}) through the kinetic energy of muon antineutrinos in Eq.~(\ref{MuAnm}). 
This term is nonlocal in the sense that $\phi$ is determined by the global distribution of the muonic charge. 
Given the total energy density (\ref{rhoTotal}), one can relate it to the pressure $P$ as per \Cref{P(rho)}.
We then have 6 equations (\ref{neutr}), (\ref{Mu_npemu}), (\ref{hydro}) for 6 functions $n_b$, $\delta$, $n_e$, $n_\mu$, $m$, $\Phi$. 
Solving them numerically with appropriate boundary conditions (see next subsection) gives the desired NS mass and density profiles.
In particular, one obtains the Mass-Radius relation shown in Fig.~\ref{fig:M_R_relation}.

\subsection{Numerical method}
\label{sec:Numerical method}

Here we describe the numerical procedure used to solve for the equation of state (E.o.S.) of the NS. As discussed above, the local parameters characterising the structure of the star, $n_b,\delta,n_e,n_\mu$, are determined by pressure $P$ and the $\LL$ potential $\phi$. In the numerical analysis it is more convenient to calculate $n_b,n_e,n_\mu,\rho,P$ as functions of $\delta$ and $g|\phi|$. 

We scan over values of $\delta$ in the range $0\leqslant\delta\leqslant 1$ with the step $0.001$ and over $g|\phi|$ in the range $0\leqslant g|\phi|\leqslant 200$ MeV with the step 1 MeV, where $\lesssim 1\%$ difference in the E.o.S. is found compared to a larger step size $0.01$ (2 MeV) for $\delta$ ($g|\phi|$), showing good convergence. Consistent solutions of baryon density $n_b$ can be found for each pair of $(\delta,g|\phi|)$, by solving the charge neutralization condition \Cref{neutr}, with proton, electron and muon number densities expressed as functions of $(\delta,n_b,g|\phi|)$ as in \Cref{Mu_npemu,Mu_np}:
\begin{equation}
    \frac{\mu_e(\delta,n_b)^{\frac{3}{2}}}{3\pi^2\hbar^3}+\frac{(\mu_\mu(\delta,n_b,g|\phi|)^2-m_\mu^2)^{3}}{3\pi^2\hbar^3} = \frac{(1-\delta)n_b}{2}.
\end{equation}
After $n_b(\delta,g|\phi|)$ is acquired, we can derive $\rho(\delta,g|\phi|)$ from Eq.~(\ref{rhoTotal}), along with $P(\delta,g|\phi|)$ from \Cref{P(rho)}. Then we interpolate the data to get the E.o.S. $P(\rho(\delta,g|\phi|),g|\phi|)$, along with other properties $n_b(P,g|\phi|)$, $n_\mu(P,g|\phi|)$, etc.

We use the initial value problem (IVP) to integrate the hydrostatic Eq.~(\ref{hydro}) and the static electric field Eq.~(\ref{staticElectric}) from $r=0$ to $r=r_\infty$, with the initial pressure at $r=0$ iterated from $3\times10^{33}{\rm erg}/{\rm cm}^3$ to $3\times10^{37}{\rm erg}/{\rm cm}^3$, with 400 steps in total. For any given initial pressure $P(0)$, we need to know the electric potential $\phi(0)$ to start the integration. 

We introduced two new functions $f_1$ and $f_2$, such that that we only need to integrate the system in regions where matter is present and that we can locally evolve the electric potential of the new force, as follows,
\begin{equation}
    \frac{\D f_1(r)}{\D r} = n_{\mu-\tau}(r)r e^{m_A r},\qquad \frac{\D f_2(r)}{\D r} = n_{\mu-\tau}(r)r e^{-m_A r},\label{f1f2}
\end{equation}
with $n_{\mu-\tau} = n_\mu-n_{\bar{\nu}_\mu}-n_{\nu_\tau}$. They relate to the electric potential as
\begin{equation}
     \frac{\phi(r)}{g} = \frac{e^{-m_A r}}{2m_A r}f_1(r)+\frac{e^{m_A r}}{2m_A r}(f_2(r_\infty)-f_2(r)) -\frac{e^{-m_A r}}{2m_A r}f_2(r_\infty).
\end{equation}
Here $f_2(r_\infty) = \phi(0)/g$ is taken at a large radius where no charge is present, $r_\infty=\;60$ km. The value $f_2(r_\infty)$ cannot be known without the charge distribution $n_{\mu-\tau}(r)$, physically related to the fact that only one value $\phi(0)$ corresponds to the correct profile of the NS system that converges to 0 at infinity. Thus it is set as the scanning parameter that will be matched by dichotomy iterations. A bigger (smaller) $f_2(r_\infty)$ will lead to a smaller (bigger) $\lim_{r\to r_\infty}f_2(r)$ from integrating Eq.~(\ref{f1f2}). We continue to iterate $f_2(r_\infty)$ until $\lim_{r\to r_\infty}f_2(r)$ matches $f_2(r_\infty)$ within $1\%$ accuracy, which in turn ensures that the boundary condition $\phi(r_\infty)\to0$ is satisfied within $1\%$. 

The IVP 
is solved with \texttt{Julia}, using an adaptive Runge–Kutta method (Tsitouras 5(4)), with relative and absolute tolerances set to $10^{-3}$ and $10^{-6}$, respectively. 
Because of the very nonlinear character of the IVP, achieving these tolerances requires a big precision in setting the value $\phi(0)$, which amounts to $10^{-66}$ for $g=10^{-16}$, $m_A=10^{-9}\,{\rm eV}$ and even higher for larger masses. Furthermore, maintaining such precision during the IVP integration requires very small integration step size. With feasible computational costs, we obtain that the solution may exhibit the unphysical behavior ($\phi(r)$ becoming slightly positive at finite $r$) for $m_A \gtrsim 1.5\times 10^{-10}\,{\rm eV}$ and $g\gtrsim10^{-18}$. 
Requiring the resulting error in the determination of the neutrino flux to not exceed $\sim 10\%$ limits the 
gauge boson mass range in Fig.~\ref{fig:constraint1} to $m_A <1.0\times 10^{-9}\,{\rm eV}$, at which the suppression effect due to $m_AR_{NS}$ being large is noticeable.

\bibliography{Refs}

\end{document}